\documentclass[12pt,a4paper]{article}
\usepackage{tikz}
\usepackage{pgf}

\usepackage{amsmath}
\usepackage{booktabs}
\usepackage{cite}
\usepackage[colorlinks=true,linkcolor=black, citecolor=black,urlcolor=black]{hyperref}

\usepackage{multirow,graphics}
\usepackage{enumerate}
\usepackage{amstext}
\usepackage{amssymb}
\usepackage{amsmath}

\usepackage{graphicx}
\usepackage{subcaption}
\usepackage{color}

\usepackage[nodisplayskipstretch]{setspace}
\usepackage{tcolorbox}
\usepackage{float}

\usepackage{wasysym}
\usepackage{ytableau}
\usepackage{textcomp}

\usetikzlibrary{decorations.pathmorphing}
\usetikzlibrary{decorations.markings}
\usetikzlibrary{intersections}
\usetikzlibrary{calc}
\usetikzlibrary{positioning}
\usetikzlibrary{3d}
\usetikzlibrary{shapes}
\usetikzlibrary{angles}
\usetikzlibrary{fit}
\usetikzlibrary{backgrounds}
\usetikzlibrary{arrows}

\tikzset{
photon/.style={decorate, decoration={snake}},
particle/.style={postaction={decorate},
    decoration={markings,mark=at position .5 with {\arrow{>}}}},
antiparticle/.style={postaction={decorate},
    decoration={markings,mark=at position .5 with {\arrow{<}}}},
gluon/.style={decorate, decoration={coil,amplitude=2pt, segment length=4pt},color=purple},
wilson/.style={color=blue, thick},
scalarZ/.style={postaction={decorate},decoration={markings, mark=at position .5 with{\arrow[scale=1]{stealth}}}},
scalarX/.style={postaction={decorate}, dashed, dash pattern = on 4pt off 2pt, dash phase = 2pt, decoration={markings, mark=at position .53 with{\arrow[scale=1]{stealth}}}},
scalarZw/.style={postaction={decorate},decoration={markings, mark=at position .75 with{\arrow[scale=1]{stealth}}}},
scalarXw/.style={postaction={decorate}, dashed, dash pattern = on 4pt off 2pt, dash phase = 2pt, decoration={markings, mark=at position .60 with{\arrow[scale=1]{stealth}}}},
frozen/.style={inner sep=0.7mm, rectangle,draw},
frozenblue/.style={rectangle, draw, fill=blue!20, inner sep=0.7mm},
norm/.style={->, draw, shorten <=2pt, shorten >=2pt},
diag/.style={->, draw, shorten <=5pt, shorten >=3pt},
every node/.style={inner sep=0.5mm}
}

\makeatletter
\tikzoption{canvas is plane}[]{\@setOxy#1}
\def\@setOxy O(#1,#2,#3)x(#4,#5,#6)y(#7,#8,#9)%
  {\def\tikz@plane@origin{\pgfpointxyz{#1}{#2}{#3}}%
   \def\tikz@plane@x{\pgfpointxyz{#4}{#5}{#6}}%
   \def\tikz@plane@y{\pgfpointxyz{#7}{#8}{#9}}%
   \tikz@canvas@is@plane
  }
\makeatother

\newcommand\be{\begin{equation}}
\newcommand\ee{\end{equation}}

\newcommand{\atwo}{
  \mathbin{
    \tikz[]{
      \draw[white] (-0.2,-0.1) rectangle (0.9,0.1);
    \coordinate (ori) at (0,0) ;
    \coordinate (des) at (0.7,0) ;
    \draw[-latex] ([yshift=0pt]ori)  to   ([yshift=0pt]des)  ;
  }
  }
}

\newcommand{\affineatwo}{
  \mathbin{
    \tikz[]{
      \draw[white] (-0.2,-0.1) rectangle (0.9,0.1);
    \coordinate (ori) at (0,0) ;
    \coordinate (des) at (0.7,0) ;
    \draw[thick,-latex,double] ([yshift=0pt]ori)  to   ([yshift=0pt]des)  ;
    
  }
  }
}

\normalsize
\makeatletter
\divide\@tempdima\baselineskip
\@tempcnta=\@tempdima
\skip\@mpfootins = \skip\footins
\makeatletter
\DeclareFontFamily{OMX}{MnSymbolE}{}
\DeclareSymbolFont{MnLargeSymbols}{OMX}{MnSymbolE}{m}{n}
\SetSymbolFont{MnLargeSymbols}{bold}{OMX}{MnSymbolE}{b}{n}
\DeclareFontShape{OMX}{MnSymbolE}{m}{n}{
    <-6>  MnSymbolE5
   <6-7>  MnSymbolE6
   <7-8>  MnSymbolE7
   <8-9>  MnSymbolE8
   <9-10> MnSymbolE9
  <10-12> MnSymbolE10
  <12->   MnSymbolE12
}{}
\DeclareFontShape{OMX}{MnSymbolE}{b}{n}{
    <-6>  MnSymbolE-Bold5
   <6-7>  MnSymbolE-Bold6
   <7-8>  MnSymbolE-Bold7
   <8-9>  MnSymbolE-Bold8
   <9-10> MnSymbolE-Bold9
  <10-12> MnSymbolE-Bold10
  <12->   MnSymbolE-Bold12
}{}

\let\llangle\@undefined
\let\rrangle\@undefined
\DeclareMathDelimiter{\llangle}{\mathopen}%
                     {MnLargeSymbols}{'164}{MnLargeSymbols}{'164}
\DeclareMathDelimiter{\rrangle}{\mathclose}%
                     {MnLargeSymbols}{'171}{MnLargeSymbols}{'171}

\renewcommand{\@dotsep}{10000}
\makeatother

\begin{document}
\numberwithin{equation}{section}
\begin{center}
\phantom{vv}

\vspace{3cm}
\bigskip

{\Large \bf Algebraic singularities of scattering amplitudes from tropical geometry}

\bigskip 
\mbox{\bf James Drummond\footnotemark, Jack Foster$^1$,  \"Omer G\"urdo\u gan\footnotemark, Chrysostomos Kalousios$^1$}%
\setcounter{footnote}{1}
\footnotetext{\, {\texttt{\{j.a.foster, j.m.drummond, c.kalousios\}@soton.ac.uk }}}
\setcounter{footnote}{2}
\footnotetext{\,\,\texttt{omer.gurdogan@maths.ox.ac.uk}}
\bigskip

{$^1$\em School of Physics \& Astronomy, University of Southampton,\\
  Highfield, Southampton, SO17 1BJ, United Kingdom.}\\[10pt]

{$^2$\em Mathematical Institute, University of Oxford,\\
Andrew Wiles Building,
Woodstock Road,
Oxford,
OX2 6GG, United Kingdom.}

\vspace{3cm} \bigskip \vspace{30pt} {\bf Abstract}
\end{center}

We address the appearance of algebraic singularities in the symbol alphabet of scattering amplitudes in the context of planar $\mathcal{N}=4$ super Yang-Mills theory. We argue that connections between cluster algebras and tropical geometry provide a natural language for postulating a finite alphabet for scattering amplitudes beyond six and seven points where the corresponding Grassmannian cluster algebras are finite. As well as generating natural finite sets of letters, the tropical fans we discuss provide letters containing square roots. Remarkably, the minimal fan we consider provides all the square root letters recently discovered in an explicit two-loop eight-point NMHV calculation.

\noindent 

\newpage
\phantom{vv}
\vspace{1cm}
\hrule
\tableofcontents

\bigskip
\medskip

\hrule
%\newpage

\section{Introduction}

Loop amplitudes in perturbative quantum field theory exhibit an intricate analytic structure. Understanding this  structure in greater depth has allowed many advances and pushed the boundaries of what is computationally feasible by an enormous amount. In addition to the obvious practical benefits of a greater understanding, there has also been a surprising interplay with modern advances in mathematics, for example in the theory of polylogarithmic functions and their elliptic counterparts. In planar $\mathcal{N}=4$ super Yang-Mills theory there are many connections with various areas of mathematics, essentially because the theory is simple enough to allow for explicit results at higher orders in perturbation theory, while still being rich enough to exhibit many different analytic features.

Of particular relevance to this article is the connection between the singularities of planar loop amplitudes in $\mathcal{N}=4$ super Yang-Mills and cluster algebras related to Grassmannian spaces ${\rm Gr}(4,n)$. Cluster algebras were introduced and developed in \cite{1021.16017,1054.17024,fomin_zelevinsky_2007} and are an area of intense ongoing research. Their relation to scattering amplitudes was first discussed in \cite{ArkaniHamed:2012nw} in the context of on-shell diagrams. The connection to the branch cut singularities of amplitudes was established in \cite{Golden:2013xva} and explored further in e.g. \cite{Golden:2014xqa}. This connection relates the cluster $\mathcal{A}$-coordinates of  the cluster algebra with the symbol letters (potential logarithmic branch cuts) of the scattering amplitude. The cluster algebra connection explains the simple nine-letter alphabet of singularities previously found in six-particle amplitudes \cite{Goncharov:2010jf} and it has been exploited in the context of the analytic bootstrap programme up to high loop orders \cite{Dixon:2011pw,Dixon:2013eka,Dixon:2014voa,Dixon:2014iba,Dixon:2015iva,Caron-Huot:2016owq,Caron-Huot:2019vjl}. Moreover the link to cluster algebras suggests a 42 letter alphabet for seven-particle amplitudes which has successfully been used to bootstrap amplitudes in \cite{Drummond:2014ffa,Dixon:2016nkn,Drummond:2018caf}. 

Further support for the underlying connection to cluster algebras in the structure of scattering amplitudes comes from the discovery of cluster adjacency \cite{Drummond:2017ssj,Drummond:2018dfd}. This is an analytic property of amplitudes which relates different singularities to each other. In particular only cluster $\mathcal{A}$-coordinates which appear together in some cluster may appear in adjacent slots of the symbol. This property implies the Steinmann relations used in \cite{Caron-Huot:2016owq,Dixon:2016nkn} and their extended versions \cite{Caron-Huot:2019bsq} (and under the assumption of physical branch cuts on the Euclidean sheet, seems also to be implied by them). An important point about the property of cluster adjacency is that it extends the role of the cluster algebra beyond the union of the $\mathcal{A}$-coordinates it generates; it also provides a role for the way the clusters themselves appear in the algebra. The property can also be phrased geometrically in terms of boundary facets of a polytope, only singularities corresponding to boundary components with the appropriate intersection can appear next to each other in the symbol. Pairs of letters corresponding to non-intersecting boundary components may not appear as neighbours.

Although the original connection to cluster algebras was inspired by the all-multiplicity result for two-loop MHV amplitudes in \cite{CaronHuot:2011ky}, it has been clear for some time that additional ingredients are needed when going beyond seven points. In the first instance the cluster algebras are finite type only for ${\rm Gr}(4,6)$ and ${\rm Gr}(4,7)$. For ${\rm Gr}(4,8)$ and beyond there are infinitely many cluster $\mathcal{A}$-coordinates, so some truncation to a finite set needs to be specified, as happens for the two-loop MHV amplitudes. Moreover at eight points and beyond there is an additional problem which is present already at one loop for N${}^2$MHV amplitudes. Four-mass box configurations appear which have letters which are not rational when expressed in terms of the Pl\"ucker coordinates for the Grassmannian spaces (i.e. in terms of momentum twistors \cite{Hodges:2009hk}). Algebraic letters were also predicted for the two-loop NMHV amplitude \cite{Prlina:2017azl,Prlina:2017tvx} by means of a Landau analysis (as initiated in this context in \cite{Dennen:2016mdk}) of the integrand provided by the amplituhedron \cite{Arkani-Hamed:2013jha,Arkani-Hamed:2017vfh}. Letters containing square roots appear in the eight-point integrals considered in \cite{Bourjaily:2018aeq,Bourjaily:2019igt}. Recently, a two-loop NMHV calculation \cite{Zhang:2019vnm} based on solving the $\overline{Q}$-equation of \cite{CaronHuot:2011kk,Bullimore:2011kg} for the dual octagonal (super) Wilson loop \cite{Alday:2007hr,Drummond:2007aua,Brandhuber:2007yx,Mason:2010yk,CaronHuot:2010ek} has revealed a specific set of 18 multiplicatively independent algebraic letters in addition to 180 rational ones.

Here we propose that an answer to both problems may be provided by tropical geometry. Recently we investigated tropical fans associated to positive Grassmannians in \cite{Drummond:2019qjk}. This investigation was in part motivated by the connection made in \cite{Cachazo:2019ngv} between tropical Grassmannians and scattering equations and their generalisations. In this context the tropical Grassmannian is related to a generalisation of tree-level biadjoint $\phi^3$ amplitudes. In the course of that investigation we highlighted the connection of the tropical geometry to the associated Grassmannian cluster algebra, a connection explored in part already in \cite{2003math.....12297S}. This connection will again play a central role in relating tropical fans to the singularities of scattering amplitudes, as we will discuss in the following.

\subsubsection*{Note added} 
Related topics on cluster algebras and scattering amplitudes are discussed in \cite{ALS,HP}.

\section{Review of positive tropical ${\rm Gr}(4,8)$}
\label{sect-2}

Following the methods described by Speyer and Williams \cite{2003math.....12297S}, in \cite{Drummond:2019qjk} we initiated a study of the fan describing the positive part of the tropical Grassmannian ${\rm Gr}(4,8)$. 
Here we will describe further features of the positive tropical Grassmannian ${\rm Gr}(4,8)$ which lead to the emergence of non-rational letters. Specifically, the ${\rm Gr}(4,8)$ cluster algebra is not finite, but of affine type $E_7^{(1,1)}$ \cite{Felikson_2012}. This feature means that although the algebra is infinite, the infinity is controlled in a particular way and it makes ${\rm Gr}(4,8)$ a very natural example to consider in going beyond the finite cases.
The affine nature of the cluster algebra leads us to natural infinite sequences of clusters which play a role in fully defining the Speyer-Williams fan (and related fans). Remarkably, the simplest infinite sequences lead to \emph{exactly} the set of non-rational letters recently discovered in the two-loop eight-point NMHV amplitude \cite{Zhang:2019vnm}.

\begin{figure}
\begin{center}
{\footnotesize
\begin{tikzpicture}
\pgfmathsetmacro{\nw}{1.3}
\pgfmathsetmacro{\vvwnw}{2.5}
\pgfmathsetmacro{\vvvwnw}{2.85}
\pgfmathsetmacro{\nh}{0.6}
\pgfmathsetmacro{\aa}{0.6}
\pgfmathsetmacro{\ep}{0.1}
\node at (-0.5*\nw -\aa,\aa+0.5*\nh) {$\langle 1\,2\,3\,4 \rangle$};
\draw[] (-\aa,\aa) -- (-\aa -\nw,\aa) -- (-\aa -\nw, \aa+\nh) -- (-\aa,\aa+\nh) -- cycle;
\node at (0.5*\nw +0*\aa,-0*\aa-0.5*\nh) {$\langle 1\,2\,3\,5 \rangle$};
\node at (0.5*\nw +0*\aa,-1*\aa-1.5*\nh) {$\langle 1\,2\,4\,5 \rangle$};
\node at (0.5*\nw +0*\aa,-2*\aa-2.5*\nh) {$\langle 1\,3\,4\,5 \rangle$};
\node at (0.5*\nw +0*\aa,-3*\aa-3.5*\nh) {$\langle 2\,3\,4\,5 \rangle$};
\draw[] (0,-3*\aa-3*\nh) -- (0,-3*\aa-4*\nh) -- (\nw,-3*\aa-4*\nh) -- (\nw,-3*\aa-3*\nh) -- cycle;
\node at (1.5*\nw +1*\aa,-0*\aa-0.5*\nh) {$\langle 1\,2\,3\,6 \rangle$};
\node at (1.5*\nw +1*\aa,-1*\aa-1.5*\nh) {$\langle 1\,2\,5\,6 \rangle$};
\node at (1.5*\nw +1*\aa,-2*\aa-2.5*\nh) {$\langle 1\,4\,5\,6 \rangle$};
\node at (1.5*\nw +1*\aa,-3*\aa-3.5*\nh) {$\langle 3\,4\,5\,6 \rangle$};
\draw[] (\nw+\aa,-3*\aa-3*\nh) -- (\nw+\aa,-3*\aa-4*\nh) -- (2*\nw+\aa,-3*\aa-4*\nh) -- (2*\nw+\aa,-3*\aa-3*\nh) -- cycle;
\node at (2.5*\nw +2*\aa,-0*\aa-0.5*\nh) {$\langle 1\,2\,3\,7 \rangle$};
\node at (2.5*\nw +2*\aa,-1*\aa-1.5*\nh) {$\langle 1\,2\,6\,7 \rangle$};
\node at (2.5*\nw +2*\aa,-2*\aa-2.5*\nh) {$\langle 1\,5\,6\,7 \rangle$};
\node at (2.5*\nw +2*\aa,-3*\aa-3.5*\nh) {$\langle 4\,5\,6\,7 \rangle$};
\draw[] (2*\nw+2*\aa,-3*\aa-3*\nh) -- (2*\nw+2*\aa,-3*\aa-4*\nh) -- (3*\nw+2*\aa,-3*\aa-4*\nh) -- (3*\nw+2*\aa,-3*\aa-3*\nh) -- cycle;
\node at (3.5*\nw +3*\aa,-0*\aa-0.5*\nh) {$\langle 1\,2\,3\,8 \rangle$};
\draw[] (3*\nw+3*\aa,-0*\aa-0*\nh) -- (3*\nw+3*\aa,-0*\aa-1*\nh) -- (4*\nw+3*\aa,-0*\aa-1*\nh) -- (4*\nw+3*\aa,-0*\aa-0*\nh) -- cycle;
\node at (3.5*\nw +3*\aa,-1*\aa-1.5*\nh) {$\langle 1\,2\,7\,8 \rangle$};
\draw[] (3*\nw+3*\aa,-1*\aa-1*\nh) -- (3*\nw+3*\aa,-1*\aa-2*\nh) -- (4*\nw+3*\aa,-1*\aa-2*\nh) -- (4*\nw+3*\aa,-1*\aa-1*\nh) -- cycle;
\node at (3.5*\nw +3*\aa,-2*\aa-2.5*\nh) {$\langle 1\,6\,7\,8 \rangle$};
\draw[] (3*\nw+3*\aa,-2*\aa-2*\nh) -- (3*\nw+3*\aa,-2*\aa-3*\nh) -- (4*\nw+3*\aa,-2*\aa-3*\nh) -- (4*\nw+3*\aa,-2*\aa-2*\nh) -- cycle;
\node at (3.5*\nw +3*\aa,-3*\aa-3.5*\nh) {$\langle 5\,6\,7\,8 \rangle$};
\draw[] (3*\nw+3*\aa,-3*\aa-3*\nh) -- (3*\nw+3*\aa,-3*\aa-4*\nh) -- (4*\nw+3*\aa,-3*\aa-4*\nh) -- (4*\nw+3*\aa,-3*\aa-3*\nh) -- cycle;
\draw[->] (-\aa+0*\ep,\aa-\ep) -- (0-0*\ep,0+\ep);
\draw[->] (0.5*\nw,-\nh-\ep) -- (0.5*\nw,-\nh-\aa+\ep);
\draw[->] (0.5*\nw,-2*\nh-\aa-\ep) -- (0.5*\nw,-2*\nh-2*\aa+\ep);
\draw[->] (0.5*\nw,-3*\nh-2*\aa-\ep) -- (0.5*\nw,-3*\nh-3*\aa+\ep);
\draw[->] (1*\nw+\ep,-0.5*\nh) -- (1*\nw+\aa-\ep,-0.5*\nh);
\draw[->] (1*\nw+\ep,-1.5*\nh-\aa) -- (1*\nw+\aa-\ep,-1.5*\nh-\aa);
\draw[->] (1*\nw+\ep,-2.5*\nh-2*\aa) -- (1*\nw+\aa-\ep,-2.5*\nh-2*\aa);
\draw[->] (1*\nw+\aa-0*\ep,-\nh-\aa+\ep) -- (1*\nw+0*\ep,-\nh-\ep);
\draw[->] (1*\nw+\aa-0*\ep,-2*\nh-2*\aa+\ep) -- (1*\nw+0*\ep,-2*\nh-1*\aa-\ep);
\draw[->] (1*\nw+\aa-0*\ep,-3*\nh-3*\aa+\ep) -- (1*\nw+0*\ep,-3*\nh-2*\aa-\ep);
\draw[->] (1.5*\nw+1*\aa,-\nh-\ep) -- (1.5*\nw+1*\aa,-\nh-\aa+\ep);
\draw[->] (1.5*\nw+1*\aa,-2*\nh-1*\aa-\ep) -- (1.5*\nw+1*\aa,-2*\nh-2*\aa+\ep);
\draw[->] (1.5*\nw+1*\aa,-3*\nh-2*\aa-\ep) -- (1.5*\nw+1*\aa,-3*\nh-3*\aa+\ep);
\draw[->] (2.5*\nw+2*\aa,-\nh-\ep) -- (2.5*\nw+2*\aa,-\nh-\aa+\ep);
\draw[->] (2.5*\nw+2*\aa,-2*\nh-\aa-\ep) -- (2.5*\nw+2*\aa,-2*\nh-2*\aa+\ep);
\draw[->] (2.5*\nw+2*\aa,-3*\nh-2*\aa-\ep) -- (2.5*\nw+2*\aa,-3*\nh-3*\aa+\ep);
\draw[->] (2*\nw+\aa+\ep,-0.5*\nh) -- (2*\nw+2*\aa-\ep,-0.5*\nh);
\draw[->] (2*\nw+\aa+\ep,-1.5*\nh-\aa) -- (2*\nw+2*\aa-\ep,-1.5*\nh-\aa);
\draw[->] (2*\nw+\aa+\ep,-2.5*\nh-2*\aa) -- (2*\nw+2*\aa-\ep,-2.5*\nh-2*\aa);
\draw[->] (2*\nw+2*\aa-0*\ep,-\nh-\aa+\ep) -- (2*\nw+\aa+0*\ep,-\nh-\ep);
\draw[->] (2*\nw+2*\aa-0*\ep,-2*\nh-2*\aa+\ep) -- (2*\nw+\aa+0*\ep,-2*\nh-1*\aa-\ep);
\draw[->] (2*\nw+2*\aa-0*\ep,-3*\nh-3*\aa+\ep) -- (2*\nw+\aa+0*\ep,-3*\nh-2*\aa-\ep);
\draw[->] (3*\nw+2*\aa+\ep,-0.5*\nh) -- (3*\nw+3*\aa-\ep,-0.5*\nh);
\draw[->] (3*\nw+2*\aa+\ep,-1.5*\nh-\aa) -- (3*\nw+3*\aa-\ep,-1.5*\nh-\aa);
\draw[->] (3*\nw+2*\aa+\ep,-2.5*\nh-2*\aa) -- (3*\nw+3*\aa-\ep,-2.5*\nh-2*\aa);
\draw[->] (3*\nw+3*\aa-0*\ep,-\nh-\aa+\ep) -- (3*\nw+2*\aa+0*\ep,-\nh-\ep);
\draw[->] (3*\nw+3*\aa-0*\ep,-2*\nh-2*\aa+\ep) -- (3*\nw+2*\aa+0*\ep,-2*\nh-1*\aa-\ep);
\draw[->] (3*\nw+3*\aa-0*\ep,-3*\nh-3*\aa+\ep) -- (3*\nw+2*\aa+0*\ep,-3*\nh-2*\aa-\ep);
\end{tikzpicture}
}
\end{center}
\caption{The initial cluster of the Grassmannian cluster algebra ${\rm Gr}(4,8)$.}
\label{Gr48initial}
\end{figure}
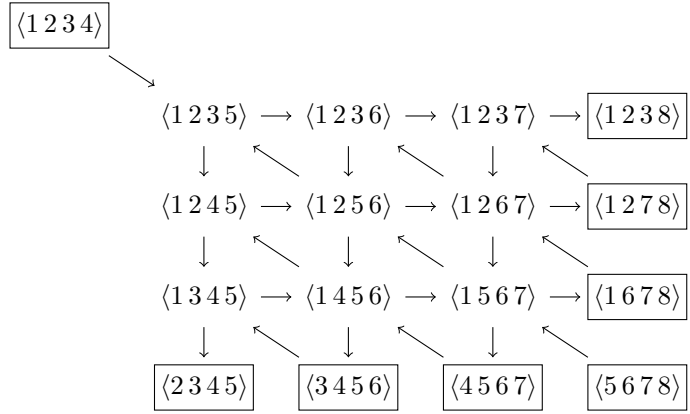

Let us recall some of the basic features of cluster algebras associated to Grassmannians \cite{scott_2006}. Our presentation will essentially follow that of \cite{fomin_zelevinsky_2007}, with the notation adapted to our conventions. A cluster algebra can be specified by some choice of initial cluster which can be encoded in a quiver diagram. The quiver diagram comprises a set of nodes, each labelled by a generator of the algebra called a cluster $\mathcal{A}$-coordinate. The nodes are either \emph{active} or \emph{frozen} and are connected by a network of arrows. The ${\rm Gr}(4,8)$ cluster algebra which is the focus of our interest here has an initial cluster of the form shown in Fig. \ref{Gr48initial} with $\mathcal{A}$-coordinates given by Pl\"ucker variables $\langle ijkl\rangle$. It has nine active nodes $a_i$ (labelled $1,\ldots,9$ from the top left and descending column by column) and eight frozen nodes $f_i$ indicated by boxes making 17 nodes in total,
\begin{align}
\!\{ a_1,..., a_9\} & =  \{\langle 1235 \rangle, \langle 1245 \rangle, \langle 1345 \rangle, \langle 1236 \rangle, \langle 1256 \rangle, \langle 1456 \rangle, \langle 1237 \rangle, \langle 1267 \rangle, \langle 1567 \rangle \},\notag \\
\!\{ f_1,..., f_8\} &  =  \{\langle 1234 \rangle, \langle 2345 \rangle, \langle 3456 \rangle, \langle 4567 \rangle, \langle 5678 \rangle, \langle 1678 \rangle, \langle 1278 \rangle, \langle 1238 \rangle \}\,.
\end{align}
When we need to consider all 17 $\mathcal{A}$-coordinates together we order them as follows: $\{a_1,...,a_9,f_1,...,f_8 \}$.

The arrows of the quiver diagram can be described by a square matrix $b$ (the \emph{exchange matrix}) with entries
\be
b_{ij} =  (\text{no. of arrows } i\rightarrow j) - (\text{no. of arrows } j \rightarrow i)\,.
\ee
Here the matrix $b$ is skew-symmetric\footnote{More generally in the study of cluster algebras it need only be skew-symmetrisable.} with indices running over all nodes (active and frozen) and in the case of ${\rm Gr}(4,8)$ therefore has dimension $(17 \times 17)$. We do not need to record arrows between frozen nodes so the bottom right $(8 \times 8)$ submatrix of $b$ is irrelevant in what follows.

In addition to the $\mathcal{A}$-coordinates and the $b$ matrix we have more data associated to the initial cluster. We also have a \emph{coefficient matrix}, taken to be the $(9 \times 9)$ identity matrix. Additionally, to each active node $a_i$ we associate the ${\bf g}$-vector ${\bf e}_i$, the unit vector in the $i$th direction.

Given the data for the initial cluster we may obtain the data for every other cluster by repeated mutation on active nodes. If we mutate on node $k$ we obtain a new $b$ matrix,
\begin{equation} \label{eq:bmutation}
b_{ij}' = \begin{cases}
	-b_{ij} & \text{if $i=k$ or $j=k$}. \\
	b_{ij} + [-b_{ik}]_+ b_{kj} + b_{ik} [b_{kj}]_+ & \text{otherwise}.
	\end{cases}
\end{equation}
where $[x]_+ = \max(x,0)$.
The $\mathcal{A}$-coordinate associated to the mutated node becomes
\begin{equation}
a_k ' = \frac{1}{a_k} \prod^{m+n}_{i=1} a_{i}^{[b_{ik}]_+} + \prod^{m+n}_{i=1} a_{i}^{[-b_{ik}]_+} .
\end{equation}
Note that, despite the denominator, the $\mathcal{A}$-coordinates can always be expressed as polynomials in Pl\"ucker coordinates after making use of Pl\"ucker relations.
The coefficient matrix also transforms as follows\footnote{Note that our conventions for the transformation of the coefficient matrix and the ${\bf g}$-vectors are modified with respect to those of Fomin and Zelevinsky \cite{fomin_zelevinsky_2007} by replacing $b \rightarrow - b$.},
\begin{equation}
c_{ij}' = \begin{cases}
	-c_{ij} & \text{if $j=k$}. \\
	c_{ij} - [-c_{ik}]_+ b_{kj} + c_{ik} [-b_{kj}]_+ & \text{otherwise}.
	\end{cases}
\end{equation}
Finally the ${\bf g}$-vector on node $k$ mutates as follows,
\begin{align}
%\mathbf{g}_l ' &= \mathbf{g}_l, \quad \text{for} \quad l \neq k \notag \\
\mathbf{g}_k ' &= -\mathbf{g}_k + \sum_{i=1}^{n} [-b_{ik}]_{+} \mathbf{g}_i + \sum_{j=1}^{n} [c_{jk}]_{+} \mathbf{b}_{j}^{0}
\label{gmut}
\end{align}
where $\mathbf{b}_{j}^{0} \text{, } j \in \{1, \ldots, 9 \} $ corresponds to the $j$th column of $b^0$, the exchange matrix for the initial cluster.
By following the above rules one may obtain every cluster in the cluster algebra. In particular to each $\mathcal{A}$-coordinate generated there will be an associated ${\bf g}$-vector. For this reason we also use the notation ${\bf g}(a)$ for the ${\bf g}$-vector associated to the $\mathcal{A}$-coordinate $a$.  As we described in \cite{Drummond:2019qjk} the ${\bf g}$-vectors play a role in describing a tropical fan associated with the positive part of the tropical Grassmannian.

To describe the tropical fan of \cite{2003math.....12297S} we first introduce the cluster $\mathcal{X}$-coordinates. These may be obtained from the $\mathcal{A}$-coordinates of some cluster by writing for each active node $j$,
\be
x_j = \prod_{i=1}^{17} a_i^{b_{ij}}\,,
\ee
where the product ranges over all $\mathcal{A}$ coordinates (active and frozen).
From the initial cluster we obtain a set of cluster $\mathcal{X}$-coordinates,
\begin{align}
&x_{11} = \frac{\langle 1234\rangle \langle 1256 \rangle}{\langle 1236\rangle \langle 1245 \rangle}  &&x_{12} = \frac{\langle 1235\rangle \langle 1267 \rangle}{\langle 1237 \rangle \langle 1256 \rangle}  &&x_{13} = \frac{\langle 1236\rangle \langle 1278 \rangle}{\langle 1238 \rangle \langle 1267 \rangle} \notag \\
&x_{21} = \frac{\langle 1235\rangle \langle 1456 \rangle}{\langle 1256\rangle \langle 1345 \rangle}  &&x_{22} = \frac{\langle 1236 \rangle \langle 1245 \rangle \langle 1567 \rangle }{\langle 1235 \rangle \langle 1456 \rangle \langle 1267 \rangle}  &&x_{23} = \frac{\langle 1237 \rangle \langle 1256\rangle \langle 1678 \rangle  }{\langle 1236 \rangle  \langle 1567 \rangle \langle 1278 \rangle} \notag \\
&x_{31} = \frac{\langle 1245\rangle \langle 3456 \rangle}{\langle 1456\rangle \langle 2345 \rangle}  &&x_{32} = \frac{\langle 1256 \rangle \langle 1345 \rangle \langle 4567 \rangle }{\langle 1245 \rangle \langle 3456 \rangle \langle 1567 \rangle}  &&x_{33} = \frac{\langle 1267 \rangle \langle 1456 \rangle \langle 5678 \rangle }{\langle 1256 \rangle \langle 4567 \rangle \langle 1678 \rangle} \,,
\label{initialX}
\end{align}
where we have chosen a labelling using a pair of indices for future convenience. This labelling is related to the usual  labelling as follows
\be
\{ x_1,\ldots,x_9\} = \{x_{11},x_{21},x_{31},x_{12},x_{22},x_{32},x_{13},x_{23},x_{33}\}\,.
\ee
We may use the $\mathcal{X}$-coordinates (\ref{initialX}) to parametrise a $(4 \times 8)$ matrix $W$ (the web matrix) of the form 
\be
W = \left( 1\!\!1_{4} | M \right)\,,
\ee
where the $(4 \times 4)$ matrix $M$ has entries $m_{ij}$ given as a sum over Young tableaux of at most $(4-i)$ rows $\underline{\lambda} = \{\lambda_1,\ldots,\lambda_{4-i}\}$ with at most $(j-1)$ columns,
\be
m_{ij} = (-1)^i \sum_{\underline{\lambda}\in Y_{ij}} \prod_{k=1}^{4-i} \prod_{l=1}^{\lambda_k} x_{kl}\,,
\ee
where $Y_{ij}$ means the range $0\leq \lambda_{4-i} \leq \ldots \leq \lambda_1 \leq j-1$. The above formula is equivalent to the sum over paths of the web diagram described in \cite{2003math.....12297S}.

The minors $\langle ijkl \rangle$, formed from the columns $i,j,k,l$ of the web matrix $W$ evaluate to polynomials in the cluster $\mathcal{X}$-coordinates (\ref{initialX}). They do so in such a way that the ratios of products of minors in (\ref{initialX}) correctly evaluate to the $\mathcal{X}$-coordinates themselves. As examples of minors we find for instance
\begin{align}
\langle 1247 \rangle &= 1 + x_{11} + x_{11} x_{12} \,,\notag\\
\langle 2346 \rangle &= 1 + x_{11} + x_{11} x_{21} + x_{11} x_{21} x_{31}\,.
\end{align}

To describe the positive tropical Grassmannian following \cite{2003math.....12297S} we evaluate these minors tropically. That is, we replace addition with minimum and multiplication with addition,
\begin{align}
{\rm Trop} \langle 1247 \rangle &= {\rm min}(0,\tilde{x}_{11}, \tilde{x}_{11} + \tilde{x}_{12}) \,,\notag\\
{\rm Trop} \langle 2346 \rangle &= {\rm min}(0,\tilde{x}_{11}, \tilde{x}_{11} + \tilde{x}_{21},  \tilde{x}_{11}+ \tilde{x}_{21} + \tilde{x}_{31})\,,
\label{tropicalminors}
\end{align}
where we remind the reader that these are tropical polynomials by using $\tilde{x}$ instead of $x$.
Each tropical minor defines a number of regions (each one a cone) of piecewise linearity in the $\tilde{x}$ space. Taking all tropical minors together we get many such regions whose overlap defines a fan. Each maximal cone of the fan is a region in which all tropical minors are linear functions. If we intersect the fan with the unit sphere in the (nine-dimensional) space of the $\tilde{x}$, each maximal cone becomes an eight-dimensional facet of a polyhedral complex.

The boundaries of the facets are locations where at least one minor is between two different regions of piecewise linearity. For example, the minor ${\rm Trop} \langle 1247 \rangle$ in (\ref{tropicalminors}) has boundaries between regions of piecewise linearity if one of the following tropical hypersurface conditions holds,
\begin{align}
&\tilde{x}_{11} = 0 \leq \tilde{x}_{11} + \tilde{x}_{12}\, \notag \\
\text{or} \quad &\tilde{x}_{11} + \tilde{x}_{21} = 0 \leq \tilde{x}_{11} \, \notag \\
\text{or} \quad &\tilde{x}_{11} = \tilde{x}_{11} + \tilde{x}_{22} \leq 0 \,.
\end{align}
Each eight-dimensional facet has seven-dimensional boundaries where one such condition is obeyed. The boundaries themselves have six-dimensional boundaries where two linearly independent equalities and the associated inequalities are obeyed. Proceeding in this way we arrive at zero-dimensional boundaries, called \emph{rays}, where eight linearly independent tropical hypersurface conditions are obeyed. 

In fact one may generalise the above discussion and consider multiple different tropical fans associated to a given Grassmannian. We could consider a fan defined by only a subset of minors, for example only those minors of the form $\langle i\, i+1 \,j\, j+1 \rangle$ or $\langle i-1\, i\, i+1\, j\rangle$. Or we could refine the fan further by including tropical evaluations of cluster $\mathcal{A}$-coordinates which are polynomials in minors, as well as the minors themselves. More generally we will define a fan by choosing some subset $\mathcal{S}$ of tropically evaluated $\mathcal{A}$-coordinates and we denote the fan by $F(\mathcal{S})$. The fan of Speyer and Williams described above then corresponds to the choice where $\mathcal{S}$ is the set of all minors.

It is important to emphasise that for any given choice of the set $\mathcal{S}$, the resulting fan is finite and in particular has a finite number of rays. One may systematically solve the tropical hypersurface conditions to find all the rays of some fan $F(\mathcal{S})$. In \cite{Drummond:2019qjk} we described another approach which makes use of the associated cluster algebra. In the case of finite cluster algebras, including the finite Grassmannian algebras studied in \cite{Drummond:2019qjk}, the cluster algebra also defines a fan by means of its ${\bf g}$-vectors.
In fact the ${\bf g}$-vector fan coincides with the fan obtained by considering $\mathcal{S}$ to be given by the set of all $\mathcal{A}$-coordinates (not just all minors). It is therefore in general a refinement of the Speyer-Williams fan. In the case of ${\rm Gr}(2,n)$ the $\mathcal{A}$-coordinates are all minors and the ${\bf g}$-vector fan coincides with the Speyer-Williams fan.

In the case of the Grassmannian ${\rm Gr}(4,8)$ we cannot immediately define a fan using all the cluster $\mathcal{A}$-coordinates since there are infinitely many. We can nevertheless use the ${\bf g}$-vectors of the cluster algebra as candidate rays of any fan $F(\mathcal{S})$ defined by tropical evaluation of a finite set $\mathcal{S}$ of cluster $\mathcal{A}$-coordinates. If we restrict ourselves to looking for rays, this approach is very effective. Systematically constructing the rays of the fan can be quite cumbersome for large fans but, given a candidate ray, it is trivial to check if it is truly a ray. As we already outlined in \cite{Drummond:2019qjk}, if we consider the Speyer-Williams fan where we take $\mathcal{S}$ to be the set of all minors then we find that 356 ${\bf g}$-vectors of the cluster algebra are also rays of the fan. 

We can similarly determine that for $\mathcal{S} = \{\langle i\,i+1\,j\,j+1\rangle\,,\, \langle i-1\,, i\,,i+1\,,j\rangle\}$ (the maximal parity-invariant subset of minors) we find  that 272 ${\bf g}$-vectors are rays. For $\mathcal{S} = \{ \langle ijkl \rangle \,,\, \langle \overline{ i j k l }\rangle\}$ (the parity completion of all minors) we find that 544 ${\bf g}$-vectors are rays. Passing from the cluster algebra to a choice of fan defined by a set $\mathcal{S}$ of $\mathcal{A}$-coordinates is therefore a natural way to obtain a finite truncation of the infinite set of cluster $\mathcal{A}$ coordinates.

Most interestingly, in none of the above cases do the ${\bf g}$-vectors provide a complete set of rays. In fact we find additional rays which complete the above sets of ${\bf g}$-vectors as shown in Table \ref{rays}. As we will describe in the next section, the cluster algebra can also be used to find the extra rays which are not ${\bf g}$-vectors. In fact they arise as limits of special infinite sequences of ${\bf g}$-vectors so we refer to them as \emph{limit rays}.
  \begin{table}[h]
    \centering
    \begin{tabular}{c|cc}
            \toprule
            $\mathcal{S}$ & g-vector rays & limit rays\\
            \midrule
            $\{\langle i\,i+1\,j\,j+1\rangle\,,\, \langle i-1\, i\,i+1\,j\rangle\}$ & 272 & 2 \\
            $\{\langle ijkl \rangle\}$ & 356 & 4 \\
            $\{ \langle ijkl \rangle \,,\, \langle \overline{ i j k l }\rangle\}$ & 544 & 4\\
            \bottomrule
\end{tabular}
\caption{Number of rays of the fans $F(\mathcal{S})$ for different choices of $\mathcal{S}$. }
\label{rays}
\end{table}

To each ${\bf g}$-vector is associated a cluster $\mathcal{A}$-coordinate. We will conclude this section by explicitly listing the $\mathcal{A}$-coordinates corresponding to the 272 ${\bf g}$-vector rays in the least refined fan described in Table \ref{rays}. If we also include the eight frozen $\mathcal{A}$-coordinates then the resulting 280 $\mathcal{A}$ coordinates contain the 196 rational letters found in \cite{Prlina:2017tvx} as an alphabet predicted by Landau analysis for the two-loop NMHV amplitude. In fact the explicit result for the two-loop octagon found recently in \cite{Zhang:2019vnm} contains only 180 of these rational letters. In addition the two-loop NMHV octagon contains 18 multiplicatively independent algebraic letters involving square roots, only four of which (corresponding to the letters of the possible four-mass box integral topologies) are contained in the list in \cite{Prlina:2017tvx}.

We begin the list of 280 letters (including 8 frozen) by recalling the 196 rational letters of \cite{Prlina:2017tvx},
\begin{itemize}
\item 68 four-brackets of the form $\left< a\, a+1\, b \,c\right>$,

\item 8 cyclic images of $\left<12\bar{4} \cap \bar{7} \right>$,

\item 40 cyclic images of $\left<1(23)(45)(78)\right>$, $\left<1(23)(56)(78)\right>$, $\left<1(28)(34)(56)\right>$, $\left<1(28)(34)(67)\right>$, $\left<1(28)(45)(67)\right>$,

\item 48 dihedral images of $\left<1(23)(45)(67)\right>$, $\left<1(23)(45)(68)\right>$, $\left<1(28)(34)(57)\right>$,

\item 8 cyclic images of $\left<\bar{2} \cap (245)\cap\bar{8}\cap(856)\right>$,

\item 8 distinct images of $\left<\bar{2} \cap (245)\cap\bar{6}\cap(681)\right>$,

\item 16 dihedral images of $\llangle 12345678 \rrangle$.
\end{itemize}
In addition, we have the following 84 rational letters,
\begin{itemize}
\item 2 letters, $\left<1357\right>$ and $\left<2468\right>$,

\item 8 cyclic images of $\left<1(23)(46)(78)\right>$ (this set is closed under reflections),

\item 16 dihedral images of $\left<1(27)(34)(56)\right>$,

\item 2 cyclic images of $\left<\bar{2}\cap\bar{4}\cap\bar{6}\cap\bar{8}\right>$ (this set returns to itself under two rotations and it is closed under reflections),

\item 8 cyclic images of $\left<\bar{2}\cap(246)\cap\bar{6}\cap\bar{8}\right>$ (this set is closed under reflections),

\item 32 dihedral images of $\llangle 12435678 \rrangle$, $\llangle 12436578 \rrangle$,

\item 16 dihedral images of $
\langle 1234 \rangle \langle 1678 \rangle \langle 2456 \rangle-
\langle 1267 \rangle \langle 1348 \rangle \langle 2456 \rangle+
\langle 1248 \rangle \langle 1267 \rangle \langle 3456 \rangle
$.
\end{itemize}
In the above we have defined $\llangle abcdefgh \rrangle = 
\langle abcd \rangle \langle abef \rangle \langle degh \rangle-
\langle abde \rangle \langle abef \rangle \langle cdgh \rangle+
\langle abde \rangle \langle abgh \rangle \langle cdef \rangle$.

In an ancillary file we list the g-vectors and their corresponding letters for the first two cases of Table 1.

We now turn to describing the extra rays obtained by limits of infinite sequences and the resulting algebraic letters.
\section{Infinite paths in ${\rm Gr}(4,8)$ and algebraic letters}

As we have seen in the previous discussion, the relation between amplitude singularities and cluster algebra data requires some refinement when going beyond seven points. In the first instance, the two-loop NMHV octagon has algebraic letters which do not correspond to any cluster $\mathcal{A}$-coordinate. In addition, in truncating the infinite set of $\mathcal{A}$-coordinates by considering some tropical fan $F(\mathcal{S})$ as described above, the rays of $F(\mathcal{S})$ are not all described by ${\bf g}$-vectors of the cluster algebra.

We may address both of the above difficulties by realising that the infinite number of clusters can usefully be organised into infinite sequences, each of which can be related to an infinite rank two cluster algebra with two nodes and a doubled arrow between them. Such algebras were considered in e.g. \cite{reading2018combinatorial} and it was already noted there that under repeated mutation the ${\bf g}$-vectors asymptote to a limiting vector. In fact, in the affine case which is relevant here, the same limiting vector can be obtained by repeated mutation with either choice of initial node (i.e. both directions asymptote to the same limit vector).

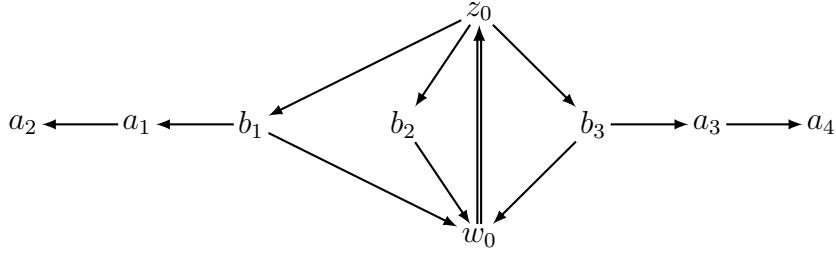
\begin{figure}
\begin{center}
  \begin{tikzpicture}[node distance=2cm]
  %\node (A)  (0,0) [circle,fill,inner sep=2pt]{} ;
  \node  (w0)  at (0,0) {$w_0$}; 
  \node  (z0)  at (0,3) {$z_0$};
  \node  (b1)  at (-3,1.5) {$b_1$};
  \node  (b2)  at (-1,1.5) {$b_2$};
  \node  (b3)  at (1.5,1.5) {$b_3$};
    \node  (a1)  at (-4.5,1.5) {$a_1$};
    \node  (a2)  at (-6,1.5) {$a_2$};
    \node  (a3)  at (3,1.5) {$a_3$};
    \node  (a4)  at (4.5,1.5) {$a_4$};
   \draw[double,thick,->=,>=latex] (w0) -- (z0) ;
   \draw[thick,->,>=latex] (b1) -- (w0) ;
    \draw[thick,->,>=latex] (b2) -- (w0) ;
     \draw[thick,->,>=latex] (b3) -- (w0) ;
   \draw[thick,->,>=latex] (z0) -- (b1) ;
    \draw[thick,->,>=latex] (z0) -- (b2) ;
     \draw[thick,->,>=latex] (z0) -- (b3) ;
      \draw[thick,->,>=latex] (b1) -- (a1) ;
      \draw[thick,->,>=latex] (a1) -- (a2) ;
      \draw[thick,->,>=latex] (b3) -- (a3) ;
      \draw[thick,->,>=latex] (a3) -- (a4) ;
    \end{tikzpicture}
     \caption{The $E_{7}^{(1,1)}$ shaped clusters with a doubled arrow between two cluster $\mathcal{A}$-coordinates, $w_0$ and $z_0$. By mutation on the $a_i$ nodes we generate an $A_2 \times A_2$ subalgebra of clusters containing the same $w_0$, $z_0$ and $b_i$ nodes. Frozen nodes are omitted here.}
   \label{E711cluster}
\end{center}

\end{figure}

If we ignore the frozen nodes (and ignore the values of the $\mathcal{A}$-coordinates on the active nodes) there are 506 distinct quiver diagrams that arise in the ${\rm Gr}(4,8)$ cluster algebra. The fact that there are only finitely many is a feature of the affine cases of Grassmannian cluster algebras ${\rm Gr}(4,8)$ and ${\rm Gr}(3,9)$ and these algebras are referred to as \emph{finite mutation type}. Out of the 506 quivers, 491 have only single arrows while 15 have a doubled arrow. These latter type have the shape of the $E_7^{(1,1)}$ quiver diagram shown in Fig. \ref{E711cluster}, or one related to it by mutation in the $A_2 \times A_2$ subalgebra generated by mutations on the $a_i$ type nodes \cite{Felikson_2012}.

\begin{figure}
\begin{center}
  \begin{tikzpicture}[node distance=2cm]
  \node  (w0)  at (0,0) {$w_0$}; 
  \node  (z0)  at (2,0) {$z_0$};
  \node  (b)  at (1,1.5) {$b$};
  \node  (f0)  at (0,-1.5) {$f_w$};
  \node  (f1)  at (2,-1.5) {$f_z$};
  %\node    at (1,.25) {\scriptsize $2$};
    \draw[thick,double,->,>=latex] (w0) -- (z0) ;
    \draw[thick,->,>=latex] (b) -- (w0) ;
    \draw[thick,->,>=latex] (z0) -- (b) ;
    \draw[thick,->,>=latex] (w0) -- (f0) ;
    \draw[thick,->,>=latex] (f1) -- (z0) ;
    
    \draw[thick,gray,dashed,->] (-1,-1) -- (-2,-2.25);
    
     \node  (z1p)  at (-4,-4) {$z_1$}; 
  \node  (z0p)  at (-2,-4) {$z_0$};
  \node  (bp)  at (-3,-2.5) {$b$};
  \node  (f0p)  at (-4,-5.5) {$f_w$};
  \node  (f1p)  at (-2,-5.5) {$f_z$};
    %\node   at (-3,-4+0.25) {\scriptsize $2$};
    \draw[thick,double,->,>=latex] (z0p) -- (z1p) ;
    \draw[thick,->,>=latex] (bp) -- (z0p) ;
    \draw[thick,->,>=latex] (z1p) -- (bp) ;
    \draw[thick,->,>=latex] (f0p) -- (z1p)  ;
    \draw[thick,->,>=latex] (f1p) -- (z0p) ;
    
      \draw[thick,gray,dashed,->] (3,-1) -- (4,-2.25); 
    
    \node  (w0m)  at (4,-4) {$w_0$}; 
  \node  (w1m)  at (6,-4) {$w_1$};
  \node  (bm)  at (5,-2.5) {$b$};
  \node  (f0m)  at (4,-5.5) {$f_w$};
  \node  (f1m)  at (6,-5.5) {$f_z$};
    %\node   at (5,-4+0.25) {\scriptsize $2$};
    \draw[thick,double,->,>=latex] (w1m) -- (w0m) ;
    \draw[thick,->,>=latex] (bm) -- (w1m) ;
    \draw[thick,->,>=latex] (w0m) -- (bm) ;
    \draw[thick,->,>=latex] (w0m) -- (f0m)  ;
    \draw[thick,->,>=latex] (w1m) -- (f1m) ;
    
    \draw[thick,gray,dashed,->] (-3,-6) -- (-3,-7);
    
    \node  (z1p2)  at (-4,-9) {$z_1$}; 
  \node  (z2p2)  at (-2,-9) {$z_2$};
  \node  (bp2)  at (-3,-7.5) {$b$};
  \node  (f0p2)  at (-4,-10.5) {$f_w$};
  \node  (f1p2)  at (-2,-10.5) {$f_z$};
    \draw[thick,double,->,>=latex] (z1p2) -- (z2p2) ;
    \draw[thick,->,>=latex] (bp2) -- (z1p2) ;
    \draw[thick,->,>=latex] (z2p2) -- (bp2) ;
    \draw[thick,->,>=latex] (f0p2) -- (z1p2)  ;
    \draw[thick,->,>=latex] (z2p2) -- (f1p2) ;
    \draw[thick,double,->,>=latex] (f1p2) -- (z1p2) ;
    
      \draw[thick,gray,dashed,->] (5,-6) -- (5,-7);
    
    \node  (w2m2)  at (4,-9) {$w_2$}; 
  \node  (w1m2)  at (6,-9) {$w_1$};
  \node  (bm2)  at (5,-7.5) {$b$};
  \node  (f0m2)  at (4,-10.5) {$f_w$};
  \node  (f1m2)  at (6,-10.5) {$f_z$};
    \draw[thick,double,->,>=latex] (w2m2) -- (w1m2) ;
    \draw[thick,->,>=latex] (bm2) -- (w2m2) ;
    \draw[thick,->,>=latex] (w1m2) -- (bm2) ;
    \draw[thick,->,>=latex] (f0m2) -- (w2m2)  ;
    \draw[thick,->,>=latex] (w1m2)  -- (f1m2) ;
     \draw[thick,double,->,>=latex] (w1m2)  -- (f0m2) ;
     
        \draw[thick,gray,dashed,->] (5,-11) -- (5,-12);
        \draw[thick,gray,dashed,->] (-3,-11) -- (-3,-12);
        
        \node  at (-3,-13) {$z_{n+2} z_n = \mathcal{C} \mathcal{F}^n + z_{n+1}^2$};
        \node  at (5,-13) {$w_{n+2} w_n = \tilde{\mathcal{C}} \mathcal{F}^n + w_{n+1}^2$};
  \end{tikzpicture}
  \caption{The doubly infinite sequence corresponding to the embeddings of the affine $A_2$ cluster algebra into the ${\rm Gr}(4,8)$ cluster algebra. After mutating one step from the origin cluster on either node, the repeated mutations give rise to a regular recurrence relation.}
   \label{doublesequence}
  \end{center}
\end{figure}
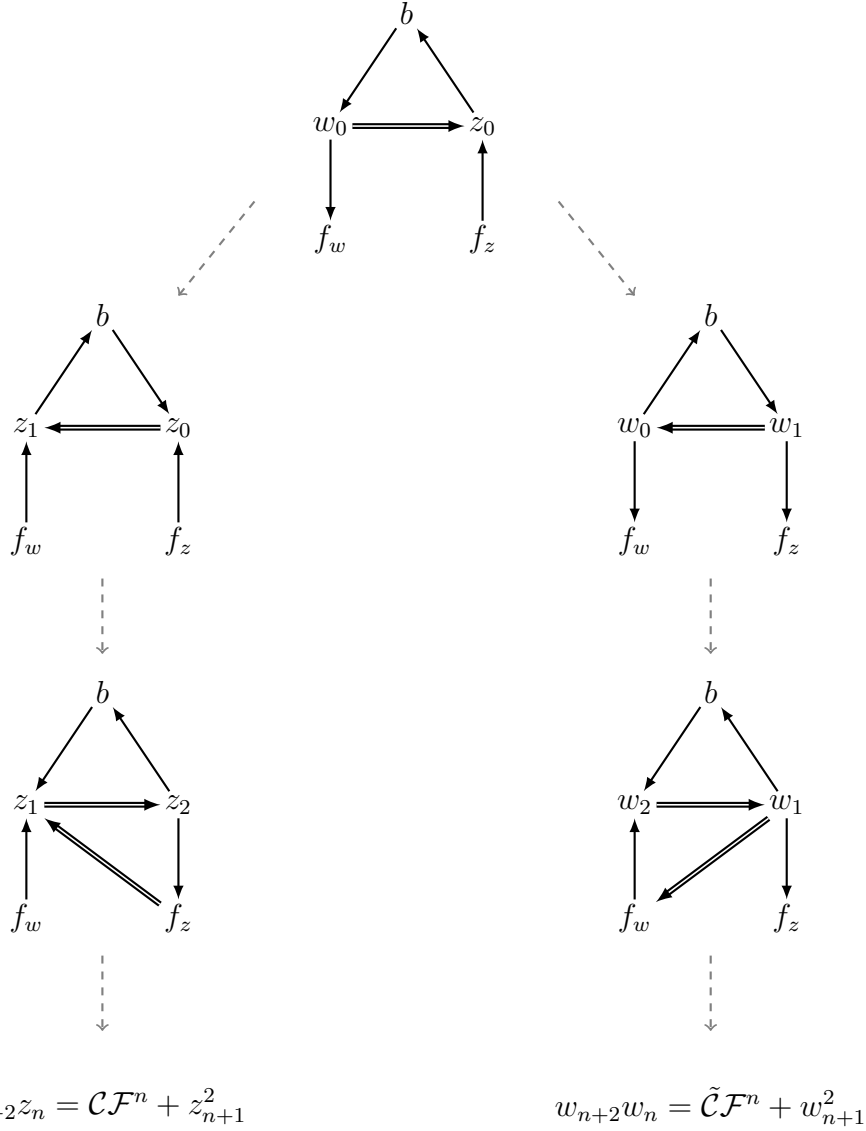

Each diagram of the form of Fig. \ref{E711cluster} forms part of a doubly infinite rank-two affine sequence, generated by alternating mutations on the $w_0$ and $z_0$ nodes. In each such sequence we can find some cluster (actually an $A_2 \times A_2$ subalgebra of clusters) in which the frozen nodes are all outgoing from $w_0$ and incoming to $z_0$. We illustrate this by a simplified diagram which we refer to as an \emph{origin cluster} where we ignore the $a_i$ nodes, combine the three $b_i$ nodes into a single node,
\be
b=b_1b_2b_3\,,
\ee
and combine all frozen nodes outgoing from $w_0$ into $f_w$ and those incoming to $z_0$ into $f_z$,
\begin{align}
f_w = \prod_{i=1}^8 f_i^{m_i}\,,\qquad f_z = \prod_{i=1}^8 f_i^{n_i}\,, \qquad m_i,n_i \in \mathbb{N}_0\,.
\end{align}
Such a simplified diagram is illustrated at the top of Fig. \ref{doublesequence}.

The initial mutations to generate the infinite double sequence take the form
\begin{align}
z_1 w_0 &= b + f_w z_0^2\,, \notag \\
w_1 z_0 &= b + f_z w_0^2 \,.
\label{firstmuts}
\end{align}
Thereafter the mutations in the $z$-direction and $w$-direction take the uniform form for $n \geq 0$,
\begin{align}
z_{n+2} z_n &= \mathcal{C} \mathcal{F}^n + z_{n+1}^2\,, \notag \\
w_{n+2} w_n &= \tilde{\mathcal{C}} \mathcal{F}^n + w_{n+1}^2\,.
\label{quadrec}
\end{align}
The coefficients $\mathcal{C}$ and $\tilde{\mathcal{C}}$ are given by
\begin{align}
\mathcal{C} &= b f_z\,, \notag \\
\tilde{\mathcal{C}} &= b f_w\,,
\end{align}
while the factor $\mathcal{F}$ is the product over the frozen nodes,
\be
\mathcal{F} = f_w f_z\,.
\ee

The transformations of the ${\bf g}$-vectors while performing the doubly infinite sequence of mutations are very simple. After a few initial mutations the differences in consecutive ${\bf g}$-vectors stabilise and we arrive at the form
\be
{\bf g}(z_{n+1}) - {\bf g}(z_n) = {\bf g}(w_0) - {\bf g}(z_0) = {\bf g}(w_{n+1}) - {\bf g}(w_n)\,.
\ee
This shows that in either direction the ${\bf g}$-vectors will asymptote to the limit ray
\be
{\bf g}_{\infty} = {\bf g}(w_0) - {\bf g}(z_0)\,.
\ee
In fact we find many different origin clusters of the form shown at the top of Fig. \ref{doublesequence}, with different $w_0$ and $z_0$ (and hence different ${\bf g}(w_0)$ and ${\bf g}(z_0)$) but with the same limit ray ${\bf g}_{\infty}$.

We may recast the quadratic recurrence relations (\ref{quadrec}) in a matrix form,
\be
\left(
\begin{matrix}
z_{n+2} & z_{n+1} \\
z_{n+1} & z_n
\end{matrix}
\right)
= 
\left(
\begin{matrix}
z_{n+1} & z_{n} \\
z_{n} & z_{n-1}
\end{matrix}
\right) 
\left(
\begin{matrix}
\mathcal{P} & 1 \\
- \mathcal{F} & 0
\end{matrix}
\right)
=
\left(
\begin{matrix}
z_{2} & z_{1} \\
z_{1} & z_0
\end{matrix}
\right)
\left(
\begin{matrix}
\mathcal{P} & 1 \\
- \mathcal{F} & 0
\end{matrix}
\right)^n\,,
\label{matrec}
\ee
and similarly for $z \rightarrow w$.
Taking the determinant of the matrix relations (\ref{matrec}) yields the original quadratic relations (\ref{quadrec}) since
\be
z_2 z_0 = \mathcal{C} + z_1^2\,.
\ee
We may verify that the matrix recursion (\ref{matrec}) consistently generates the same sequence $z_n$ as the quadratic recursion (\ref{quadrec}) provided that $\mathcal{P}$ obeys
\be
z_2 = z_1 \mathcal{P} - z_0 \mathcal{F}\,.
\label{Pz2rel}
\ee
Hence we require that $\mathcal{P}$ is related to $\mathcal{C}$ via
\be
\label{PCrel}
\mathcal{C} + z_1^2 + z_0^2 \mathcal{F} = z_0 z_1 \mathcal{P}\,.
\ee
Remarkably, $\mathcal{P}$ can be shown to be a polynomial, that is we can find a factor of $z_0 z_1$ within the combination on the LHS of (\ref{PCrel}). If we write $\mathcal{C}$ and $\mathcal{F}$ in terms of the cluster $\mathcal{A}$-coordinates of the origin cluster we find
\begin{align}
\mathcal{C} + z_1^2 + z_0^2 \mathcal{F} &= b f_z + z_1^2 + z_0^2 f_w f_z\,, \notag \\
&=  z_1 (f_z w_0 + z_1)\,,
\end{align}
where the second step is achieved by using the first relation in (\ref{firstmuts}) to eliminate $b$. We have made the factor of $z_1$ manifest and it remains to show that there is also a factor of $z_0$ in the remaining combination $(f_z w_0 + z_1)$. To show this we consider instead the square of this combination,
\begin{align}
(f_z w_0 + z_1)^2 &= f_z^2 w_0^2 + 2 f_z w_0 z_1 + z_1^2\,, \notag \\
&= f_z(w_1 z_0 - b) + 2 f_z (b+ f_w z_0^2) + (z_2 z_0 - b f_z)\,,\notag  \\
&= z_0(f_z w_1 + 2 f_z f_w z_0 + z_2)\,.
\end{align}
In the second step we have used the relations (\ref{firstmuts}) and the quadratic recurrence formula (\ref{quadrec}) for $z$ in the case $n=2$. We have succeeded in finding a factor of $z_0$ in the square factor, but since all quantities involved are \emph{polynomials} in Pl\"ucker coordinates, it must be that the original factor without the square also has a factor of $z_0$. Hence we conclude that 
\be
\mathcal{P} = \frac{f_z w_0 + z_1}{z_0}\,
\ee
is a polynomial even if this property is not manifest from the above equation. By considering the $w_n$ sequence instead we arrive at an equivalent formula for $\mathcal{P}$,
\be
\mathcal{P} = \frac{f_w z_0 + w_1}{w_0}\,.
\ee
Note that both $\mathcal{P}$ and $\mathcal{F}$ are invariant under swapping the $z$ sequence and the $w$ sequence (along with swapping $f_z$ with $f_w$). Note also that $\mathcal{P}$ is manifestly positive in the region where all $\mathcal{A}$-coordinates are positive.

Returning to the matrix recursion we see that it is equivalent to a linear recursion formula
\be
z_{n+2} = z_{n+1} \mathcal{P} - z_n \mathcal{F}\,,
\ee
of which (\ref{Pz2rel}) is just the first case. Of course we also have the same recursion formula for the $w_n$. Once a polynomial form for $\mathcal{P}$ is obtained, this linear recursion formula provides a manifestly polynomial form for all the $z_n$ cluster coordinates (and similarly the $w_n$). Note that the linear recursion would just be the Fibonacci recursion relation if we had $\mathcal{P}=-\mathcal{F}=1$.
The linear recursion formula is neatly solved by the following generating function
\be
G_z(x) = \frac{z_1 - z_0 \mathcal{F} x}{1 - \mathcal{P} x + \mathcal{F} x^2} = \sum_{n=0}^\infty z_{n+1} x^n\,,
\ee
and similarly for $w \leftrightarrow z$.
It follows immediately that the asymptotic limit of the ratios of the $z_n$ is controlled by the roots of the quadratic in the denominator,
\be
{\rm lim}_{n \rightarrow \infty} \frac{z_n}{z_{n-1}} = \mathcal{P} + \sqrt{\Delta}\,, \qquad \Delta = \mathcal{P}^2 - 4 \mathcal{F}\,.
\label{limratio}
\ee
Using this fact we can write an explicit form for the $z_n$,
\be
z_n = \frac{1}{2^{n+1}}\bigl[(z_0 + B_z \sqrt{\Delta})(\mathcal{P} + \sqrt{\Delta})^n + (z_0 - B_z \sqrt{\Delta})(\mathcal{P} - \sqrt{\Delta})^n\bigr]\,
\label{eq:sequencesolution}
\ee
with $B_z$ defined by
\be
B_z = \frac{2z_1 -z_0 \mathcal{P}}{\Delta}\,.
\ee
We have a similar formula for the $w_n$ sequence obtained by swapping $z \leftrightarrow w$ everywhere.
For a sequence of mutations generating ${\bf g}$-vectors which asymptote to a given limit ray ${\bf g}_\infty$, we find that $\mathcal{P}$ and $\mathcal{F}$ (and hence the limit of the ratio (\ref{limratio})) depend only on the limit ray. The actual path towards the limit (and therefore the $z_n$ or $w_n$) is distinguished by the values of $z_0$ and $z_1$ (or $w_0$ and $w_1$).

In the limit of large $n$, the term with $(\mathcal{P} + \sqrt{\Delta})^n$ dominates over the term $(\mathcal{P} - \sqrt{\Delta})^n$. Its coefficient $(z_0 + B_z \sqrt{\Delta})$ depends on the path of approach to the limit. Since the product $(z_0 + B_z \sqrt{\Delta})(z_0 - B_z \sqrt{\Delta})$ is rational\footnote{For the cases we consider shortly, it is always a multiplicative combination of the 280 rational letters given in at the end of Sect. \ref{sect-2}.}  we identify the ratio 
\be
\phi_z = \frac{z_0 + B_z \sqrt{\Delta}}{z_0 - B_z \sqrt{\Delta}} = \frac{z_0 \mathcal{P} - 2 f_z w_0 + z_0 \sqrt{\Delta}}{z_0 \mathcal{P} -2 f_z w_0 - z_0 \sqrt{\Delta}} 
\label{phiz}
\ee
with a new algebraic letter associated to the path. We also have a letter obtained from the limit of the $w$ sequence whose formula is the same except for swapping $z \leftrightarrow w$ everywhere,
\be
\phi_w = \frac{w_0 + B_w \sqrt{\Delta}}{w_0 - B_w \sqrt{\Delta}} = \frac{w_0 \mathcal{P} - 2 f_w z_0 + w_0 \sqrt{\Delta}}{w_0 \mathcal{P} - 2 f_w z_0 - w_0 \sqrt{\Delta}} \,.
\label{phiw}
\ee
Note that we have many origin clusters, each of which provides two paths (the $z$ branch and the $w$ branch) towards the same limit ray ${\bf g}_\infty$. The square root $\sqrt{\Delta}$ which appears will be common for all algebraic letters coming from a given limit. Only the rational coefficients (determined by the data of the origin cluster) will depend on the actual path.

Let us recall that the smallest fan from those listed in Table \ref{rays} has two limit rays in addition to the 272 ${\bf g}$-vector rays. For the case of a path that asymptotes to the first limit ray we find
\begin{align}
{\bf g}^{(1)}_\infty &= (1, -1, 0, -1, 0, 1, 0, 1, -1) \,, \notag \\
\mathcal{P} &= \langle1256\rangle\langle3478\rangle - \langle1278\rangle\langle3456\rangle - \langle1234\rangle\langle5678\rangle\,, \notag \\
\mathcal{F} &= \langle1234\rangle\langle3456\rangle\langle5678\rangle\langle1278\rangle\,. 
\label{g1lim}
\end{align}
while the second limit ray
\begin{align}
{\bf g}^{(2)}_\infty = (0, 1, 0, 1, 0, -1, 0, -1, 0) \,, 
%\notag \\
%\mathcal{P} &= - \langle2367\rangle\langle1458\rangle + \langle1238\rangle\langle4567\rangle + \langle2345\rangle\langle1678\rangle\,, \notag \\
%\mathcal{F} &= \langle2345\rangle\langle4567\rangle\langle1678\rangle\langle1238\rangle\,.
\end{align}
has $\mathcal{P}$ and $\mathcal{F}$ related to those in (\ref{g1lim}) by a cyclic rotation by one unit.
The precise values of $z_0$ and $z_1$ (or $w_0$ and $w_1$) depend on the path of approach.

We find 64 origin clusters whose associated limit rays are either ${\bf g}^{(1)}_\infty$ or ${\bf g}^{(2)}_\infty$ described above. Among them are four clusters with the nodes $w_0$ and $z_0$ connected by the doubled arrow given by
\begin{equation}
       \langle j(12)(i k)(78)\rangle
  \affineatwo
   \langle12ij\rangle\label{eq:originseed}
\end{equation}
where $i\in \{3,4\}$ and $\left(j,k\right)$ is a permutation of
$\left(5,6\right)$. Each origin cluster with the rank two affine subalgebras of the form (\ref{eq:originseed}) has the limit ray ${\bf g}^{(1)}_\infty$ and frozen nodes given by
\begin{equation}
%{  \setlength{\jot}{12pt}
%  \begin{gathered}
  f_z=f_1 = \langle 1234 \rangle ,\quad f_w=f_3f_5f_7 = \langle 3456 \rangle \langle 5678 \rangle \langle 1278 \rangle\,,
  %\\
%  \mathcal P = (1237) (8456) + (1238) ( 4567) -  (1247) ( 8356) - (1248) (3567)\, .
%\end{gathered}
%}
\end{equation}
in agreement with eq. (\ref{g1lim}). The four origin clusters are listed in Table \ref{tab:fourseeds} along with the data from the original cluster diagram that they come from, including the $b$ nodes and the $A_2 \times A_2$ subalgebra generated by the $a_i$ type nodes of Fig. \ref{E711cluster}. The full set of 64 origin clusters whose limit rays are ${\bf g}^{(1)}_\infty$ or ${\bf g}^{(2)}_\infty$ are then obtained from the four described in (\ref{eq:originseed}) by dihedral transformations.

  \begin{table}
    \centering
    \begin{tabular}{l|l|l}
            \toprule
Sub affine $A_2$: $w_0 \affineatwo z_0$ &$b=b_1 b_2 b_3$& \begin{minipage}{4cm}Residual $A_2\times A_2$\end{minipage}\\
      \midrule

                 \begin{minipage}{5.5cm}$\langle 5(12)(36)(87)\rangle \affineatwo \langle 1235 \rangle$\end{minipage}&
                     \begin{minipage}[c][2cm][c]{3.5cm}
                       $
                       \begin{array}{l}
                         \langle 1256 \rangle \\
                         \times \langle 3(12)(56)(78) \rangle \\
                         \times \langle 5(12)(34)(78) \rangle
                       \end{array}
                       $
                     \end{minipage}&
                                                                          \begin{minipage}[c][2cm][c]{3cm}
                                       $
                                       \begin{array}{l}
                                         \langle 1345 \rangle \atwo \langle 1346 \rangle \\
                                         \langle 1237 \rangle \atwo \langle 1247 \rangle
                                       \end{array}
                                       $
                                     \end{minipage}
                                          \\
            $\langle 6(12)(35)(78) \rangle \affineatwo \langle 1236 \rangle$
      &
                             \begin{minipage}[c][1.5cm][c]{3cm}
                               $
                               \begin{array}{l}
                                 \langle 1256 \rangle\\
                                 \times \langle 3(12)(56)(78) \rangle\\
                                 \times \langle 6(12)(34)(78) \rangle
                               \end{array}
                               $
                             \end{minipage}&
                                             
                                                                                  \begin{minipage}[c][2cm][c]{3cm}
                                       $
                                       \begin{array}{l}
                                         \langle 1345 \rangle \atwo \langle 1346 \rangle\\
                                         \langle 1237 \rangle \atwo \langle 3567 \rangle
                                       \end{array}
                                       $
                                     \end{minipage}
                                          \\
      $\langle 5(12)(46)(87) \rangle \affineatwo \langle 1245 \rangle$
&
                     \begin{minipage}[c][2cm][c]{3cm}
                       $
                       \begin{array}{l}
                         \langle 1256 \rangle \\
                         \times \langle 4(12)(56)(78) \rangle\\
                         \times \langle 5(12)(34)(78) \rangle
                       \end{array}
                       $
                     \end{minipage}&
                                            
                                                                          \begin{minipage}[c][2cm][c]{3cm}
                                       $
                                       \begin{array}{l}
                                         \langle 1237 \rangle \atwo \langle 1247 \rangle \\
                                         \langle 1237 \rangle \atwo \langle 1247 \rangle
                                       \end{array}
                                       $
                                     \end{minipage}
                                          \\
      $\langle 6(12)(45)(78) \rangle \affineatwo \langle 1246 \rangle$&
                             \begin{minipage}[c][1.5cm][c]{3cm}
                               $
                               \begin{array}{l}
                                 \langle 1256 \rangle \\
                                 \times \langle 4(12)(56)(78) \rangle \\
                                 \times \langle 6(12)(34)(78) \rangle
                               \end{array}
                               $
                             \end{minipage}&
                                             
                                                                                  \begin{minipage}[c][2cm][c]{3cm}
                                       $
                                       \begin{array}{l}
                                         \langle 1237 \rangle \atwo \langle 3567 \rangle \\
                                         \langle 1237 \rangle \atwo \langle 3567 \rangle
                                       \end{array}
                                       $
                                     \end{minipage}
                                          \\

      \bottomrule

\end{tabular}

\caption{Four types of clusters that act as origins of doubly-infinite
  sequences.   }
\label{tab:fourseeds}
\end{table}

Each origin cluster produces two algebraic letters $\phi_z$ and $\phi_w$ defined by eqs. (\ref{phiz}) and (\ref{phiw}). Thus we have a set of 128 algebraic letters associated to the two limit rays ${\bf g}^{(1)}_\infty$ and ${\bf g}^{(2)}_\infty$. Each limit ray is therefore associated with significantly more data than any ${\bf g}$-vector ray, each of which is associated to a single rational letter. The 128 letters associated to ${\bf g}^{(1)}_\infty$ and ${\bf g}^{(2)}_\infty$ are not all multiplicatively independent and remarkably they generate the same space as the 18 multiplicatively independent algebraic letters found in \cite{Zhang:2019vnm}! The two-loop NHMV eight-point amplitude is therefore consistent with the data obtained from the smallest fan in Table \ref{rays} in that the associated alphabet is covered by the rays of the fan.

The set of 128 algebraic letters described above is closed under parity, as the doubly infinite sequences themselves map to each other under parity. The origin clusters themselves do not necessarily map to origin clusters but sometimes map to an adjacent cluster in the infinite sequence.  In an ancillary file we explicitly list the 128 algebraic letters.

The other fans in Table \ref{rays} have four limit rays. These are similarly associated to their own set of origin clusters, again 64 such clusters, each generating two algebraic letters according to (\ref{phiz}) and (\ref{phiw}). In this case the $\mathcal{P}$ and $\mathcal{F}$ associated to ${\bf g}^{(3)}_\infty$ are as follows,
\begin{align}
{\bf g}^{(3)}_\infty = \quad &(-1, 0, 1, 0, 2, -1, 1, -1, -1) \,, \notag \\
\mathcal{P} = \quad  &\langle 1237 \rangle \langle 1458 \rangle \langle 2468 \rangle \langle 3567 \rangle
-\langle 1238 \rangle \langle 1567 \rangle \langle 2468 \rangle \langle 3457 \rangle \notag \\
- &\langle 1238 \rangle \langle 1678 \rangle \langle 2345 \rangle \langle 4567 \rangle 
 - \langle 1237 \rangle \langle 1358 \rangle \langle 2468 \rangle \langle 4567 \rangle \notag \\
- &\langle 1234 \rangle \langle 1278 \rangle \langle 3456 \rangle \langle 5678 \rangle\,, \notag \\
\mathcal{F} = \quad & \langle 1234 \rangle \langle 2345 \rangle \langle 3456 \rangle \langle 4567 \rangle \langle 5678 \rangle \langle 1678 \rangle \langle 1278 \rangle \langle 1238 \rangle \,.
\label{g3limitray}
\end{align}
The $\mathcal{P}$ associated to the other limit ray
\be
{\bf g}^{(4)}_\infty = (1, 1, -1, 1, -2, 0, -1, 0, 1)\,,
\ee
is related to that in (\ref{g3limitray}) by a cyclic rotation by one unit while the $\mathcal{F}$ is the same (note that $\mathcal{F}$ in (\ref{g3limitray}) is the product of all frozen $\mathcal{A}$-coordinates and therefore is cyclic invariant). The algebraic letters associated to the limit rays ${\bf g}^{(3)}_\infty$ and ${\bf g}^{(4)}_\infty$ are therefore of a different nature with different square roots. So far we do not have any example of an amplitude where they appear. They might appear at higher loop orders in eight-point amplitudes than are currently known explicitly.

We should also stress that there are more origin clusters (infinitely many) each of which has its own limit vector associated to it and its own type of square roots. However the limit vectors obtained are not rays of any of the fans listed in Table \ref{rays}. One could imagine making yet more refined fans $F(\mathcal{S})$ by taking yet larger sets $\mathcal{S}$ of $\mathcal{A}$-coordinates to define them. It is possible that the other limit vectors beyond the four described above become rays of such fans.

\section{Conclusions}

The fact that we find exactly the same letters appearing in \cite{Zhang:2019vnm} from tropical geometry and cluster algebras is very exciting. Ultimately we must remember that the tropical problems we have been considering here arise purely from kinematics. Momentum twistors provide a natural unconstrained set of coordinates for the kinematical space of colour-ordered amplitudes in the planar limit and dual conformal symmetry \cite{Drummond:2008vq} dictates that we should consider $sl_4$ invariant combinations of them. This leads directly to the association of the Grassmannian ${\rm Gr}(4,n)$, or more precisely ${\rm Conf}_n(\mathbb{P}^3) = {\rm Gr}(4,n)/(\mathbb{C}^*)^{n-1}$, to the kinematic space of massless scattering in planar $\mathcal{N}=4$ super Yang-Mills theory. The dual conformally invariant (or $sl_4$ invariant) quantities are the Pl\"ucker coordinates $\langle ijkl \rangle$ and they obey quadratic Pl\"ucker relations, for example the following, 
\be
\langle ijk[l \rangle \langle mnpq]\rangle = 0\,.
\ee
Tropicalising such polynomial relations gives the tropical Grassmannian as considered by Speyer and Sturmfels \cite{SpeyerSturmfels}. Considering its positive part leads to the tropical fans of Speyer and Williams \cite{2003math.....12297S}. As we have discussed, these have a direct connection to the Grassmannian cluster algebras. The ${\bf g}$-vectors of the cluster algebras provide rays for the tropical fans, even in the case where the algebra is not finite, such as the case studied here ${\rm Gr}(4,8)$. To these rays are associated rational $\mathcal{A}$-coordinates which play the role of symbol letters characterising the singularities of the polylogarithmic functions describing the scattering amplitudes. Moreover the additional rays of the fan arise as limits of natural infinite sequences of ${\bf g}$-vectors. To these are associated sets of algebraic letters involving square roots.

It remains to clarify which fans correspond to which amplitudes. We have seen that the letters of the two-loop NMHV octagon are included in the smallest fan we considered in Table \ref{rays}. However it could be that beyond two loops also the MHV amplitude will need recourse to the same set of algebraic letters. It could also be that beyond two loops the NMHV amplitude will require a bigger set of letters, say those arising in the largest fan considered in Table \ref{rays}. Moreover, the N${}^2$MHV amplitude requires algebraic letters (the four-mass box letters) at one loop already. These four algebraic letters are included in the set of 18 multiplicatively algebraic letters found in \cite{Zhang:2019vnm}. It would be very interesting to explore all these amplitudes at higher loop orders than are currently known explicitly to understand the general structure better. In a companion paper \cite{ustoappear} we investigate different fans for finite Grassmannian cluster algebras, in particular the case of ${\rm Gr}(4,7)$ where we discuss the relation of MHV and NMHV amplitudes to different possible choices of tropical fan.

\section*{Acknowledgments}
We are grateful to Nima Arkani-Hamed and Mark Spradlin for discussions about the topics described here.
JD, JF and CK are supported by ERC grant 648630 IQFT. This project
has received funding from the European Research Council (ERC) under
the European Union’s Horizon 2020 research and innovation programme
(grant agreement No. 724638)

\Urlmuskip=0mu plus 1mu\relax
\def\UrlBreaks{\do\/\do-}
\bibliographystyle{JHEP}
\bibliography{biblio}

\end{document}